\documentclass[a4paper, amsfonts, amssymb, amsmath, reprint, showkeys, nofootinbib, twoside]{revtex4-1}
\usepackage[english]{babel}
\usepackage[utf8]{inputenc}
\usepackage[colorinlistoftodos, color=green!40, prependcaption]{todonotes}
\usepackage{amsthm}
\usepackage{mathtools}
\usepackage{physics}
\usepackage{xcolor}
\usepackage{graphicx}
\usepackage[left=23mm,right=13mm,top=35mm,columnsep=15pt]{geometry} 
\usepackage{adjustbox}
\usepackage{placeins}
\usepackage[T1]{fontenc}
\usepackage{lipsum}
\usepackage{csquotes}
\usepackage[fleqn]{nccmath}
\usepackage{amssymb}
\usepackage{natbib}

\usepackage[LGRgreek]{mathastext} 

\usepackage{amsmath}
\usepackage{lineno}
\usepackage{setspace}
 \usepackage{hyperref}

\begin{document}
\title{First-principles study on phase stability and physical properties of B-site ordered Nd$_2$CrFeO$_6$ double perovskite}

\author{M. D. I. Bhuyan,\textit{$^{a}$} 
 Rana Hossain,$^{\ast}$\textit{$^{b}$} Ferdous Ara \textit{$^{c}$} and M. A. Basith$^\textit{$^{a}$}$}
    \email[Email address: \\]{mabasith@phy.buet.ac.bd \\ hossain.rana@tsme.me.es.osaka-u.ac.jp}
    \affiliation{$^{a}$Nanotechnology Research Laboratory, Department of Physics, Bangladesh University of Engineering and Technology, Dhaka 1000, Bangladesh. \\ \textit{$^{b}$Department of Mechanical Science and Bioengineering, Osaka University, Osaka 560-8531, Japan,} \\ \textit{$^{c}$Institute of Multidisciplinary Research of Advanced Materials, Tohoku University, 2-1-1, Katahira, Aoba-ku, Sendai 980-0877, Japan} \\ DOI: \href{https://doi.org/10.1039/D1CP03523A}{10.1039/D1CP03523A}}


\begin{abstract}
Here, the first-principles predictions on the structural stability, magnetic behavior and electronic structure of B-site ordered double perovskite Nd$_2$CrFeO$_6$ have been reported. Initially, the ground state of the parent single perovskites NdCrO$_3$ and NdFeO$_3$ have been studied to determine the relevant Hubbard U parameter to investigate the properties of Nd$_2$CrFeO$_6$. The thermodynamic, mechanical, and dynamic stability analyses suggest the possibility of the synthesis of Nd$_2$CrFeO$_6$ double perovskite at ambient pressure. The compound shows ferrimagnetic (FiM) nature with 2 $\mu_B$ net magnetic moment and the magnetic ordering temperature has been estimated to be $\sim$265 K. Electronic structure indicates higher probability of direct photon transition over the indirect transition with a band gap of $\sim$1.85 eV. Additional effect of Nd (4f) spin and spin-orbit coupling (SOC) on the band edges have been found to be negligible for this \textit{4f-3d-3d} spin system.  This first-principles investigation predicts that due to the ferrimagnetic nature and significantly lower band gap compared to its antiferromagnetic parent single perovskites, B-site ordered Nd$_2$CrFeO$_6$ double perovskite could be a promising material for spintronic and visible-light driven energy applications.
\end{abstract}


\maketitle

\section{Introduction}

\begin{figure*}[ht]
\centering
\includegraphics[width=17 cm]{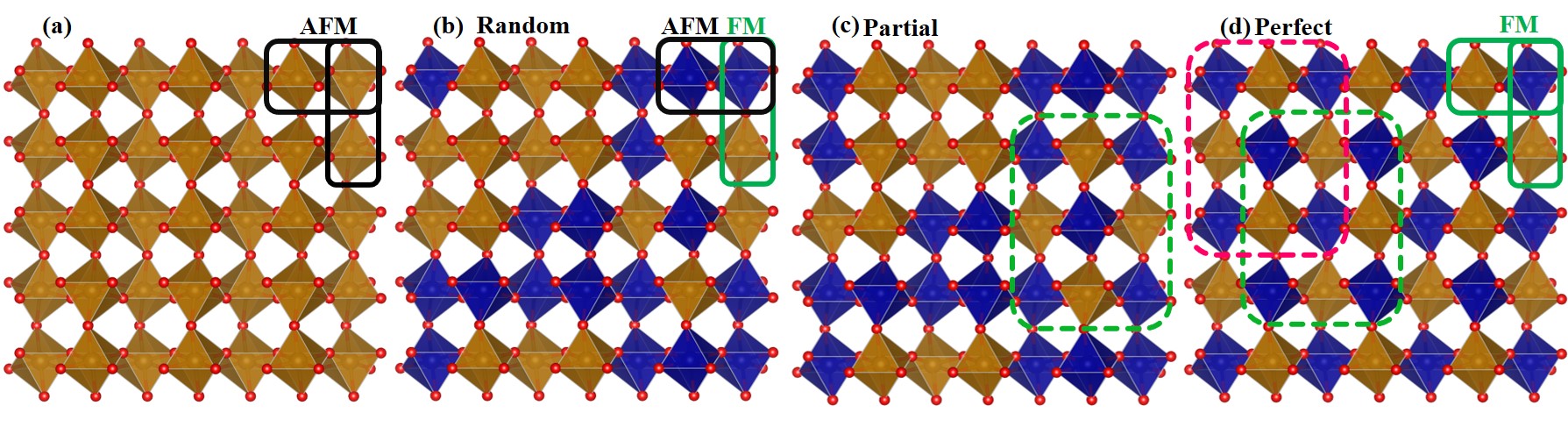}
\caption{\label{fig_1:} (a) Cation order in AB$^{'}$O$_3$ perovskite. (b) Random distribution of B$^{'}$ and B$^{''}$ cations in disordered AB$^{'}_{0.5}$B$^{''}_{0.5}$O$_3$ perovskite. (c) Partial ordering of B$^{'}$ and B$^{''}$ cations in disordered AB$^{'}_{0.5}$B$^{''}_{0.5}$O$_3$ perovskite. (d) Perfectly B$^{'}$ and B$^{''}$ sites ordered double perovskite A$_2$B$^{'}$B$^{''}$O$_6$.}
\end{figure*}

The potential of B-site ordered double perovskites A$_2$B$^{'}$B$^{''}$O$_6$ (A: alkaline or rare-earth ions \textit{e.g.}, Ca, Sr, La, Nd, Eu, Gd, Dy, \textit{etc}.; B$^{'}$, B$^{''}$: transition metals like Cr, Mn, Fe, Co, Ni, \textit{etc}.) over AB$^{'}$O$_3$ and AB$^{''}$O$_3$ single perovskites for magnetic and semiconductor-based applications have long been recognized. \cite{VASALA20151, 102133, D0CP06447E,PhysRevMaterials.2.055401,bhuyan2021sol} Double perovskites provide a wide range of structural flexibility as well as freedom of choosing a suitable combination of B$^{'}$ and B$^{''}$ cations from the periodic table  and provides opportunity to control their electronic structure, magnetic behavior, \textit{etc}. for practical application. \cite{Kim2019} To realize the opportunities of A$_2$B$^{'}$B$^{''}$O$_6$, the major challenge is to achieve the long-range B-site (B$^{'}$ and B$^{''}$ sites) ordering, which is governed by kinetic and thermodynamic factors of order-disorder reactions during synthesis of these materials.\cite{cm050064e,cm030409y} Fig. 1 shows a schematic illustration of possible B$^{'}$ and B$^{''}$ sites distribution when B$^{''}$ is inserted into AB$^{'}$O$_3$ in order to form A$_2$B$^{'}$B$^{''}$O$_6$ double perovskite. For a certain material, if the synthesis conditions are not optimized for a certain material, it may form a randomly distributed B$^{'}$ and B$^{''}$ phase or a partially ordered phase. The physical properties especially magnetism is strongly dependent on the degree of cations ordering in double perovskite. \cite{PhysRevB.99.235129} For example, ReB$^{'}_{0.5}$B$^{''}_{0.5}$O$_3$ perovskites, containing rare-earth (Re) elements, have orthorhombic (\textit{Pnma}) crystal structure similar to ReB$^{'}$O$_3$, but demonstrate enhanced magnetization because of the additional \textit{B$^{'}$-O-B$^{''}$} net ferromagnetic (FM) exchange interaction along with \textit{B$^{'}$-O-B$^{'}$}, \textit{B$^{''}$-O-B$^{''}$}, \textit{Re-O-B$^{'}$/B$^{''}$}, and \textit{Re-O-Re} antiferromagnetic (AFM) interactions.\cite{gray2010local} These materials are generally weak FMs due to the competing AFM and FM superexchange interactions arising from the B-site disordering as shown in Fig. 1(b). \cite{C7DT00032D, PhysRevB.99.235129, PhysRevB.98.134417, 1.4870139, D1DT00438G, D0CP06447E} Yuan \textit{et al.}\cite{C3NJ01046E} studied several orthorhombic ReFe$_{0.5}$Cr$_{0.5}$O$_3$ systems and reported that magnetization improves greatly upon degree of B-site ordering of Fe and Cr atoms \textit{via} the Fe$^{3+}$(d$^5$)–O–Cr$^{3+}$(d$^3$) interaction. Interestingly, due to the presence of multiple magnetic interactions, multiple magnetic phase transitions with variation in temperature have been observed corresponding to the activation of each interaction. It is also reported \cite{doi:10.1063/1.4862665} that the DyFe$_{0.5}$Cr$_{0.5}$O$_3$ perovskite shows an anomaly in temperature dependent magnetization curve near 261 K due to the Fe$^{3+}$(d$^5$)–O–Cr$^{3+}$(d$^3$) interaction and another at 120 K, related to the Cr$^{3+}$(d$^3$)–O–Cr$^{3+}$(d$^3$) interaction which is expected for DyCrO$_3$. Indeed, DyFe$_{0.5}$Cr$_{0.5}$O$_3$ also show another anomaly at 640 K corresponding to the onset of Fe$^{3+}$(d$^5$)–O–Fe$^{3+}$(d$^5$) interaction, which also observed for DyFeO$_3$.\cite{1.4870139} Similar multiple magnetic phase transitions have been reported for NdFe$_{0.5}$Cr$_{0.5}$O$_3$ disordered perovskite.\cite{SHARANNIA2017109, jpcc.1c00562} From these reports it is evident that ReB$^{'}_{0.5}$B$^{''}_{0.5}$O$_3$ retains the properties of their parent compounds and the strength of \textit{B$^{'}$-O-B$^{''}$} interaction is much weaker than that of \textit{B$^{'}$-O-B$^{'}$} and \textit{B$^{''}$-O-B$^{''}$} AFM interaction.\cite{RODRIGUES2020109334, Coutinho2018, VADURAI2015924} For the technological applications, long range \textit{B$^{'}$-O-B$^{''}$} interaction is more demanded. \cite{HOSSAIN2018414} Therefore, to achieve a very high degree of ordering in B-sites, a number of synthesis techniques like solid-state reaction, soft chemical-based synthesis routes, and thin-film deposition techniques have been proposed.\cite{cm050064e,cm030409y,1.4812368, Kleibeuker2017, PhysRevB.99.235129, PhysRevB.67.014401} Notably, in B-site ordered double perovskites, one may find three patterns (rock salt, columnar and layered) for B$^{'}$ and B$^{''}$ depending on the choice of cations and synthesis condition. \cite{B926757C, C7QI00686A} Rock salt ordering is most symmetric and common as the array of B$^{'}$ and B$^{''}$ cations is equivalent to the anion and cation positions as shown in Fig. 1(d).\cite{B926757C, C5RA08995F}

It is also noteworthy that among the double perovskites, rare-earth (Re) containing magnetic semiconductors Re$_2$B$^{'}$B$^{''}$O$_6$ have gained significant research interest because of their wide range of technological applications and rich physical properties. \cite{5.0031196, C9QI00512A} For instance, Re$_2$NiMnO$_6$ systems have been studied extensively due to their ferromagnetism, low band gap, large magneto-capacitance, magneto-resistance, and relaxor ferroelectricity.\cite{acsaelm.8b00062, NEENULEKSHMI2014285} Notably, Re$_2$CrFeO$_6$ is another class of double perovskites which is expected to be a hub of ferrimagnetic semiconductors\cite{SUN2020166670}, however, they  are less studied either by experimental or computational investigation. Double perovskite La$_2$CrFeO$_6$ is an example of this class that has been synthesized using the pulsed laser deposition technique with a high degree of Cr and Fe ordering \cite{1.4821795, PhysRevB.84.064436} and the electronic properties of this compound have been studied by the first-principles calculations.\cite{PhysRevB.85.224404} Unlike the randomly distributed Cr and Fe disordered AFM perovskite LaCr$_{0.5}$Fe$_{0.5}$O$_3$, the ordered La$_2$CrFeO$_6$ double perovskite demonstrated a ferrimagnetic (FiM) nature with a saturation magnetization of $\sim$ 2$\mu_B$ per formula unit (f.u.).\cite{1.4821795, PhysRevB.84.064436} Recently Majumder \textit{et al.}\cite{D0CP06447E} reported a saturation magnetization of $\sim0.8$ $\mu$B f.u.$^{-1}$ in partially ordered Pr$_2$FeCrO$_6$ bulk sample. Undoubtedly, a complete B-site ordered phase is difficult to form in double perovskites by conventional solid-state or sol-gel techniques. Interestingly, a perfect ordering was achieved \cite{BOOTH20091559} in several Re$_2$NiMnO$_6$ double perovskites by repeated sintering where the number of required sintering steps and sintering temperature varied with different Re cations. Therefore, extensive investigation is required to optimize different steps for the synthesis of B-site ordered Re$_2$CrFeO$_6$ double perovskites. Moreover, the experimental investigation of the material properties of these double perovskites at the electronic scale are unattainable due to the unavailability of the required experimental facilities. However, these limitations can be overcome significantly by executing the density function theory (DFT) based first-principles calculation.\cite{Yu2020, D1CP02666F}

In this investigation, the physical properties of B-site, ordered Nd$_2$CrFeO$_6$ double perovskite have been studied by generalized gradient approximation (GGA)+\textit{U} calculations. The parent single perovskites NdCrO$_3$ and NdFeO$_3$ were studied initially to understand the effect of Hubbard parameter \textit{U} on magnetic and electronic properties. Our calculations infer that if the experimental properties (\textit{e.g.} band gap and/or atomic magnetic moments) of the parent single perovskites are known, then the desired properties of the double perovskites might be investigated by optimizing the \textit{U} values of the parent compounds. This strategy could be used to understand the properties of other Re$_2$B$^{'}$B$^{''}$O$_6$ double perovskites using computationally cheaper GGA+\textit{U} calculations. Our calculation demonstrated that double perovskite Nd$_2$CrFeO$_6$ satisfied the thermodynamic, mechanical, and dynamic criteria as a stable compound. This compound showed ferrimagnetism and significantly lower band gap compared to its parent single perovskites NdCrO$_3$ and NdFeO$_3$.

\section{Computational details}

The Vienna ab-initio simulation software (VASP),\cite{KRESSE199615, PhysRevB.54.11169, PhysRevB.49.14251} was used to carry out all the calculations by using projector-augmented wave (PAW) approach. \cite{PhysRevB.50.17953, PhysRevB.59.1758} The generalized gradient approximation (GGA)\cite{PhysRevLett.77.3865, PhysRevLett.78.1396} was adopted to evaluate the electronic exchange-correlation energy using the improved Perdew–Burke–Ernzerhof (PBE) functional. Since, the GGA approach underestimates the band gap of insulators and semiconductors, the on-site Coulomb interaction was introduced using spin-polarized GGA+\textit{U} approach (Dudarev \textit{et al.}\cite{PhysRevB.57.1505}) to properly describe the localized d electrons of Fe and Cr atoms. According to the Dudarev's approach, the total energy functional is of the form
\begin{ceqn}
\begin{align*}
\small
E_{LDA(GGA)+U}=E_{LDA(GGA)}+\frac{U-J}{2}\sum_\sigma{Tr\rho^\sigma-Tr(\rho^\sigma\rho^\sigma),}
\end{align*}
\end{ceqn}
\noindent where $\rho^\sigma$ is the density matrix of d(f)-state with spin $\sigma$.\cite{PhysRevB.57.1505, doi:10.1002/qua.24521} Since the on-site Coulomb interaction \textit{U} and exchange interaction $J$ parameters are not considered separately and only the difference ($U-J$) is meaningful in Dudarev’s approach, one single parameter \textit{U} (with $J$=0) was considered in this study for simplicity.\cite{doi:10.1063} For all calculations, Fe’s 3s$^2$3p$^6$3d$^6$4s$^2$, Cr’s 3s$^2$3p$^6$3d$^5$4s$^1$ and O’s 2s$^2$2p$^4$ electrons were considered as valence electrons. Because partly filled f-state are not adequately represented by current DFT approaches and often fail to converge, the 4f electrons of the rare earth Nd atom were taken into account as core electrons in structural relaxation. \cite{PhysRevB.102.144432,Zhao2014,SUN2020166670} However, Nd's 4f electrons were considered as valence electrons to investigate the effect of 4f spin on the electronic structure, and spin–orbit coupling (SOC). Integration over the Brillouin-zone was performed in a $6\times6\times4$, $\Gamma$-centered Monkhorst–Pack $k$-point mesh. A plane-wave cutoff was set at 520 eV, and the self-consistent convergence criteria for energy was taken to 10$^{-8}$ eV. The structure was completely relaxed until the forces fell below 1 meV {\AA}$^{-1}$.

To investigate the thermodynamic stability of Nd$_2$CrFeO$_6$, the total energies of the other stable phases bounded in Nd-Fe-Cr-O phase diagram have been considered. Initial structures of the stable phases (as shown in Fig. 4) are taken from Materials Project database.\cite{Jain2013} Structural relaxation was carried out for all the phases (with \textit{U}=3.5 eV for the phases containing Fe/Cr). Phase diagram analysis has been done by using the phase diagram module the Python Materials Genomic (pymatgen) package. \cite{ong2013python, ong2008li} The symmetry of the optimized structures has been calculated by spglib package. \cite{togo2018texttt} The transition dipole moments have been studied by Vaspkit. \cite{WANG2021108033} The phonon dispersion was calculated by the Phonopy code \cite{togo2015first} using a $2\times2\times1$ supercell with a $2\times2\times2$ k-mesh.

\section{Effect of $U$ on the ground-state}

\begin{figure}[htb]
\centering
\includegraphics[width=8 cm]{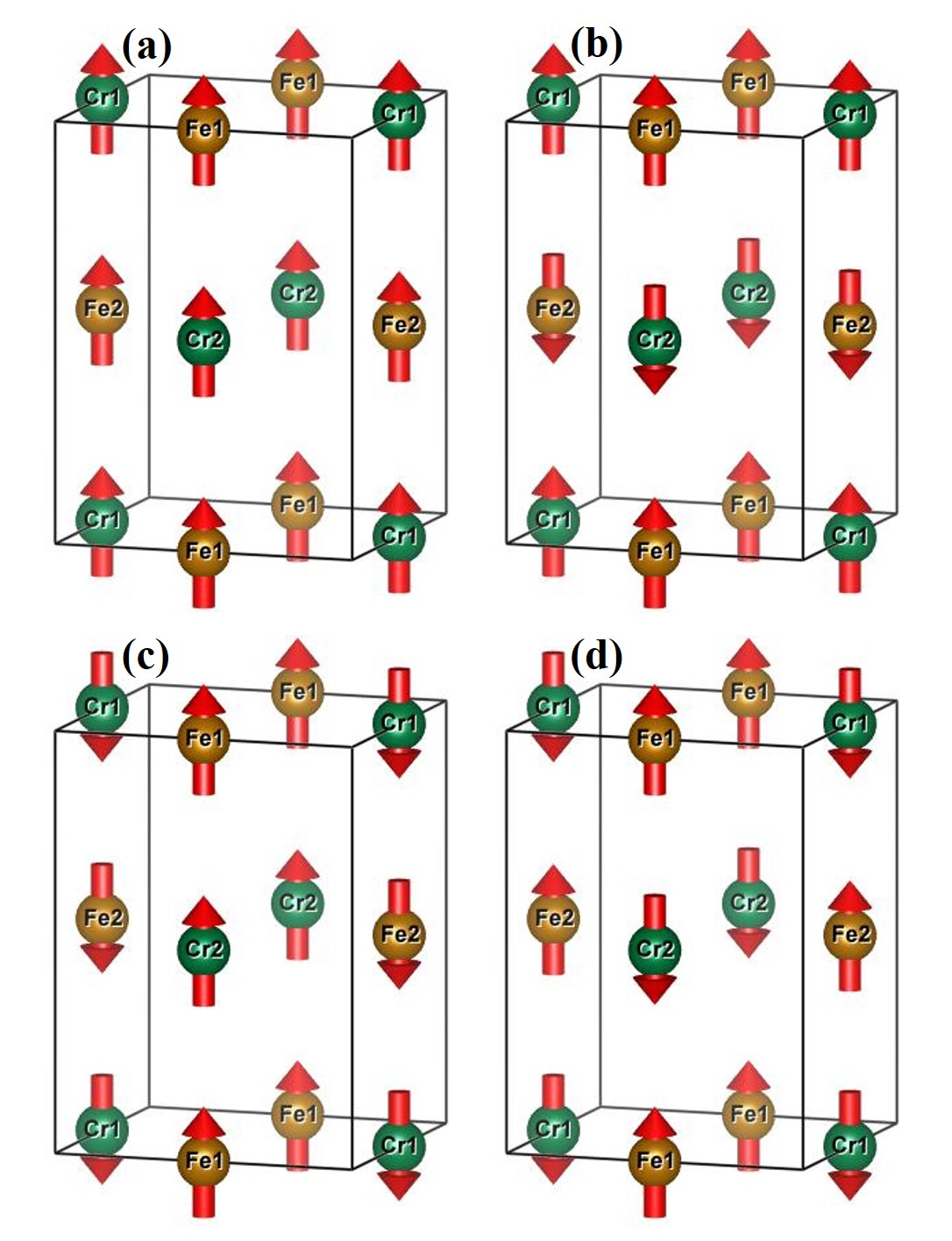}
\caption{\label{fig:} Possible spin arrangements of Fe and Cr atoms in B-site ordered Nd$_2$CrFeO$_6$ double perovskite: (a) FM, (b) A-type AFM (A-AFM), (c) C-type AFM (C-AFM), and (d) G-type AFM (G-AFM). In A-AFM orientation, the intra-plane coupling is FM while inter-plane coupling is AFM, whereas in C-AFM, this is opposite i.e., the intra-plane coupling is AFM while inter-plane coupling is FM and in G-AFM, both intra-plane and inter-plane coupling are AFM.}
\end{figure}

\begin{figure*}[htb]
 \centering
 \includegraphics[height=10cm]{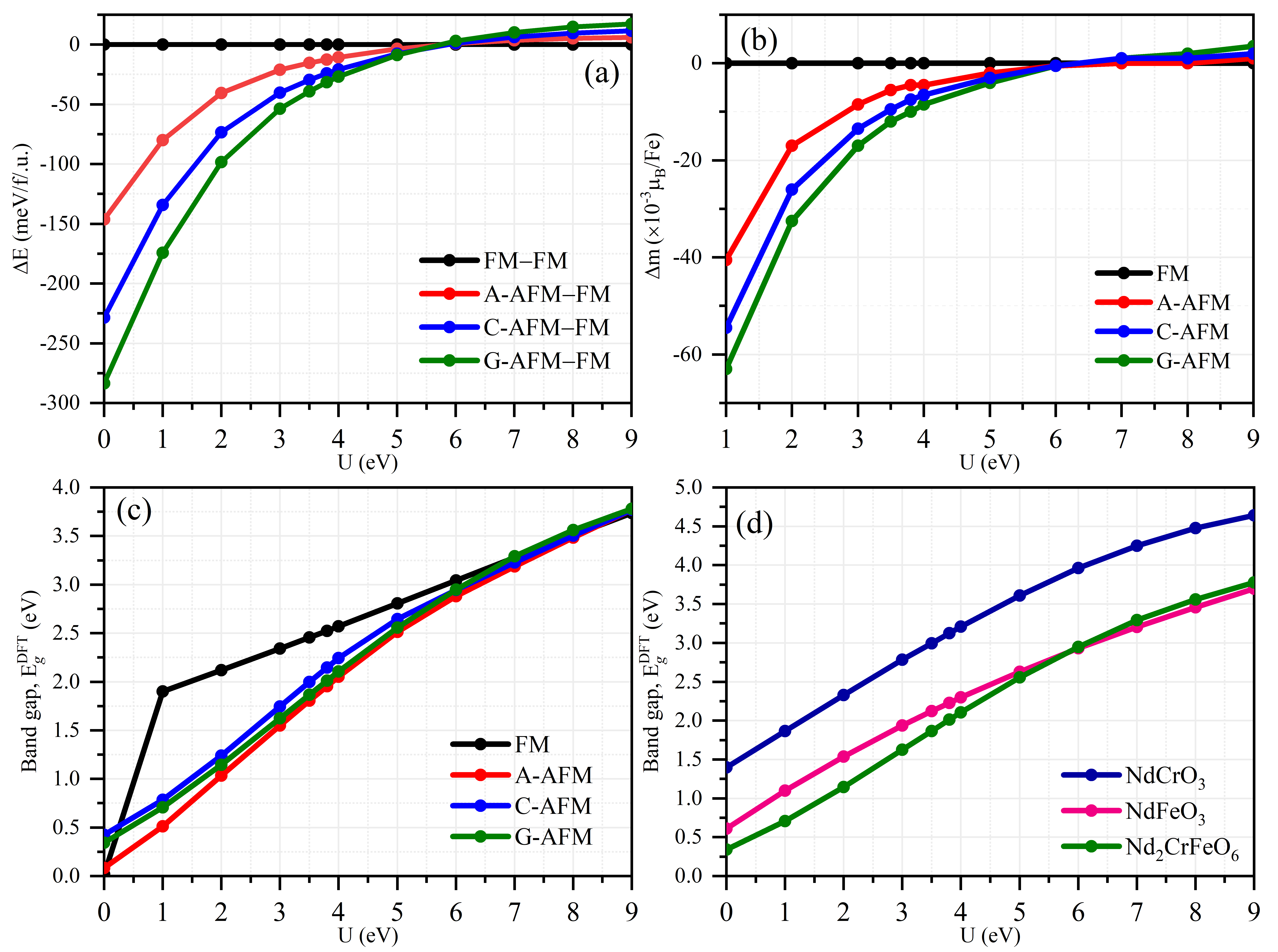}
 \caption{Effect of \textit{U} parameter on different ground-state properties. (a) Variation of ground-state energy as a function of \textit{U} for possible spin orders in Nd$_2$CrFeO$_6$ compared to the FM order. (b) Differences in spin magnetic moment (\text{$\Delta$}m) of Fe site in AFM spin orders compared to the FM order. (c) Calculated band gap (for spin down channel) for different spin orders considered for Nd$_2$CrFeO$_6$. (d) Comparative plot of band gap for NdCrO$_3$, NdFeO$_3$ single perovskite, and Nd$_2$CrFeO$_6$ double perovskite considering their low energy spin (G-AFM) state.}
 \label{fgr:example2col}
\end{figure*}

To determine the lowest energy of spin order, the FM, A-AFM, C-AFM, and G-AFM arrangements of spin (shown in Fig. 2) were considered. The structure and atomic positions were fully relaxed for a set of \textit{U} ranging from 0--9 eV for 3d electrons of Fe and Cr atoms. Fig. 3 (a) shows the total energy difference, \text{$\Delta$E} (meV f.u.$^{-1}$) of the ground states of Nd$_2$CrFeO$_6$ for various spin orders (as shown in Fig. 2) compared to the FM order and the variation corresponding to the change in \textit{U} values. The total energy of all spin orders increases with \textit{U} as the on-site occupation number in d orbital (and spin magnetic moment) of Cr/Fe found to be increased, which is consistent with Dudarev’s formalism.\cite{PhysRevB.57.1505} It should be noted that Fe/Cr spin magnetic moments increase while average induced magnetic moment in O atoms decrease from 0.09 $\mu_B$ to 0.07 $\mu_B$ with increasing \textit{U} indicating the attenuation of the strength of the magnetic exchange coupling \textit{via} Fe-O-Cr interactions for larger \textit{U}.\cite{C8RA08507B} However, \text{$\Delta$E} between FM and AFM spin orders decreases as \textit{U} increases. For \textit{U}<6 eV, G-AFM arrangement of Cr and Fe spins are energetically favorable while for larger \textit{U} values (>6 eV) FM order become the lowest energy state. The mechanism could be understood from the variation of the difference in spin magnetic moments as shown in Fig. 3(b). The variation and turning of \text{$\Delta$m} curves are identical to the \text{$\Delta$E} curves for AFM orders. It can be clearly seen that spin magnetic moments of AFM orders are approaching towards the FM baseline with increasing \textit{U} and overtaking for \textit{U}>6 eV. A similar variation has been observed for Cr. The calculated band gap for FM and AFM spin orientations are shown in Fig. 3(c) and increases smoothly with increment in \textit{U} except for FM order at \textit{U}=0 eV. At \textit{U}=0 eV, the FM spin arrangement yields metallic behavior. This is the well-known artifact of GGA in determining electronic gap of semiconductors. Prior to investigate the physical properties of Nd$_2$CrFeO$_6$ double perovskite, it is essential to choose a physically reasonable value of \textit{U} which can adequately describe the electronic structure. Therefore, the parent single perovskites were studied, since the experimental band gaps of these magnetic semiconductors already have been reported.

For single perovskites NdCrO$_3$ and NdFeO$_3$, the G-AFM arrangement of spins shows minimum energy for all values of \textit{U} with orthorhombic (\textit{Pnma}) structure and are in agreement with experimental observations. \cite{Bora2015, CHAKRABORTY201481, YAMAGUCHI1974479} Since GGA (\textit{U}=0) underestimates the band gap, the experimental band gaps (E$_g^{exp.}$) were compared with the calculated band gaps for different \textit{U} values. \cite{DAS20212408} Fig. 3(d) shows the variation in calculated band gap of NdCrO$_3$, NdFeO$_3$, and Nd$_2$CrFeO$_6$. It is found that for \textit{U}=3.50 eV (for Cr/Fe), the calculated band gaps for NdCrO$_3$ and NdFeO$_3$ matched well with experimental values.\cite{Mannepalli2017,nano11040937, 1.4954842} For the same value, \textit{i.e.} \textit{U}$_{Cr}$=\textit{U}$_{Fe}$=3.50 eV, a band gap of 1.82 eV has been obtained for double perovskite Nd$_2$CrFeO$_6$. Table 1 shows the calculated lattice parameters, symmetry of the optimized structures as well as the band gap values obtained by spin-polarized GGA and GGA+\textit{U} calculations. From the Fig. 3(d) and Table 1, it may be inferred that the band gap of double perovskite Nd$_2$CrFeO$_6$ (E$_g^{DFT}$=1.82 eV) is lower than its parent single perovskites NdCrO$_3$ (E$_g^{DFT}$=3.0 eV) and NdFeO$_3$ (E$_g^{DFT}$=2.12 eV). To understand the reason of the low band gap in Nd$_2$CrFeO$_6$, valence band maxima (VBM), conduction band minima (CBM), and the charge transfer (CT) from Cr/Fe cations to O anion have been calculated and shown in Table 1. Compared to NdFeO$_3$, the CT for Fe is found to increase in case of Nd$_2$CrFeO$_6$, while for Cr it is similar to that of NdCrO$_3$. The higher ionicity of Fe may push the Fe-3d states in the CBM to a lower-energy position together with the almost unshifted O-2p states in VBM, resulting in the band gap reduction. A previous investigation \cite{quattropani2018band} on Bi$_2$FeCrO$_6$ (band gap 1.5 eV) double perovskite also demonstrated band gap reduction from their parent single perovskites following similar mechanism. It should be noted that, the experimental band gaps of the synthesized materials depend on many factors \textit{e.g.} synthesis condition, particle size, presence of vacancies, secondary phases, \textit{etc}.,\cite{PhysRevB.99.024104, 1.4813539, ma12091444} therefore, the experimental band gap owing to these effects could vary from the DFT-calculated results for perfect system. \cite{wu2020structural} In the subsequent investigation on Nd$_2$CrFeO$_6$ double perovskite, we have used G-AFM spin orientation as it gives the lowest energy ground state and \textit{U}=3.50 eV for the 3d electrons of Cr and Fe atoms since this value can satisfactorily describe the electronic structure of the parent perovskite compounds.

\begin{table}[htb]
\centering
\small
  \caption{\ GGA calculated lattice parameters and band gaps for lowest energy  (G-AFM) state of NdCrO$_3$, NdFeO$_3$ and Nd$_2$CrFeO$_6$. Parenthesis is showing the results for (\textit{U}$_{Cr}$=\textit{U}$_{Fe}$=3.50 eV) calculations. GGA+\textit{U}-calculated position of VBM, CBM, and CT values for Cr/Fe cations to the oxygen anions obtained by Bader charge analysis.\cite{henkelman2006fast}} 
  \label{tbl:example}
  \resizebox{0.45\textwidth}{!}{\begin{tabular*}{0.48\textwidth}{@{\extracolsep{\fill}}llllll}
    \hline
     & NdCrO$_3$ & NdFeO$_3$ & Nd$_2$CrFeO$_6$ \\
    \hline
    a & 5.561 (5.586) & 5.664 (5.676) & 5.439 (5.465) \\
    b & 7.724 (7.770) & 7.797 (7.824) & 5.615 (5.627) \\
    c & 5.417 (5.449) & 5.455 (5.472) & 7.754 (7.802) \\
    angle ($\beta$) & 90 (90) & 90 (90) & 90.006 (90.012)\\
    space-group & Pnma & Pnma & P2$_1$/n  \\
    \hline
    E$_g^{DFT}$ (eV) & 1.40 (3.00) & 0.643 (2.12) & 0.33 (1.82)\\
    E$_g^{exp.}$ (eV) & 3.10\cite{Mannepalli2017} & 2.06\cite{nano11040937} & --\\   
    \hline
    VBM (eV) & 8.30 & 6.90 & 6.94 \\
    CBM (eV) & 5.30 & 4.78 & 5.12 \\    
    CT (eV/atom) & Cr=+1.73 & Fe=+1.67 & Cr=+1.73, Fe=+1.70\\  
    \hline
  \end{tabular*}}
\end{table}

\section{Phase stability}

\begin{figure}[htb]
\centering
\includegraphics[width=8.3cm]{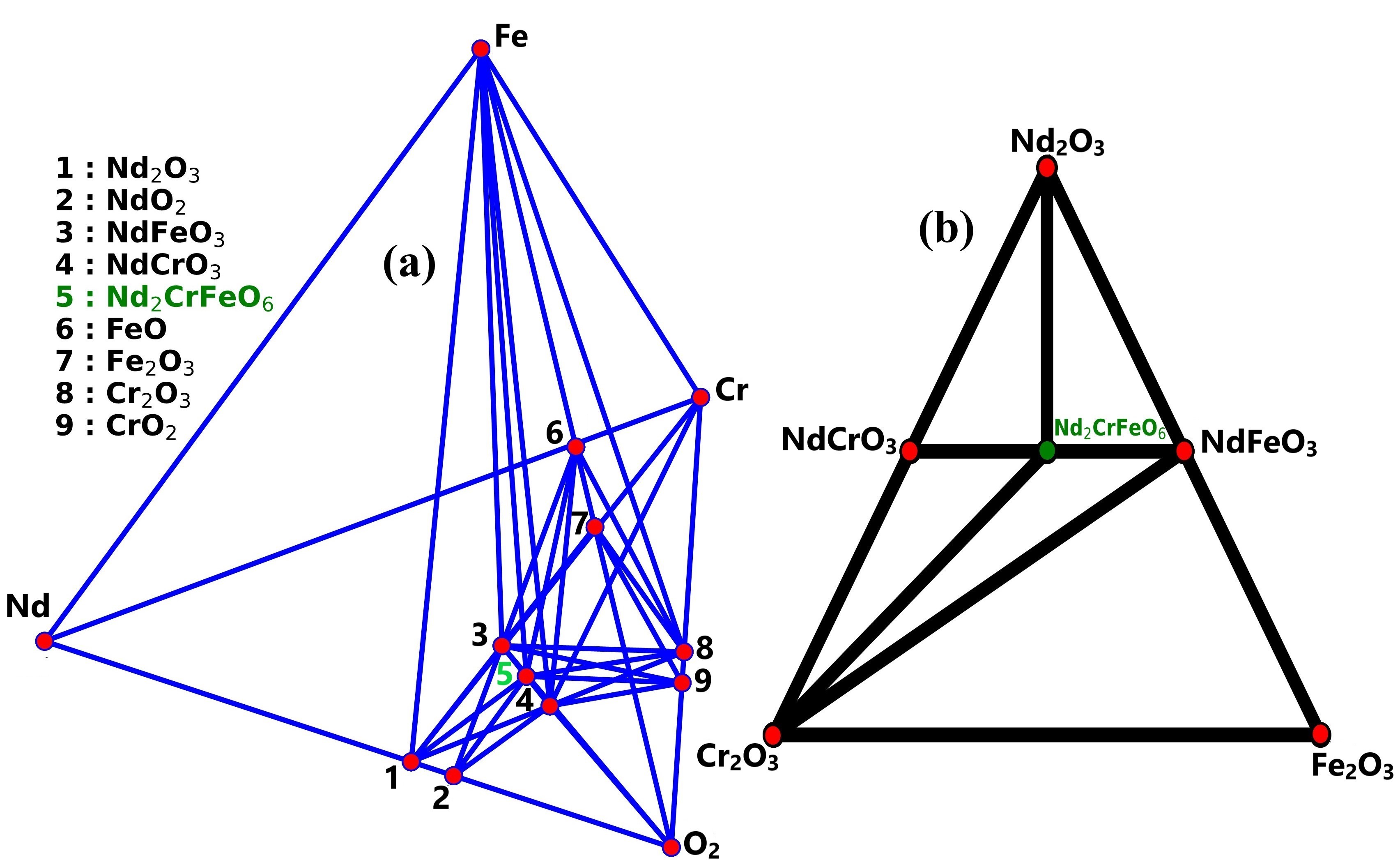}
\caption{\label{fig:epsart} (a) The DFT-calculated phase diagram for the Nd--Cr--Fe--O system at T=0 K. The red circles are indicating the stable phases, these are connected by the blue tie-lines that forming the convex hull. (b) The pseudo-ternary compound phase diagram of Nd$_2$O$_3$--Cr$_2$O$_3$--Fe$_2$O$_3$ systems, where the black tie-lines that forming the convex hull.}
\end{figure}

Fig. 4(a) and 4(b) show the constructed phase diagram for Nd-Cr-Fe-O systems and the compound phase diagram for Nd$_2$O$_3$--Cr$_2$O$_3$--Fe$_2$O$_3$ systems, respectively. It can be observed that the Nd$_2$CrFeO$_6$ phase is on the tie-line of the convex hull indicating this phase is thermodynamically stable.
Further, the formation energy per atom of Nd$_2$CrFeO$_6$ has been calculated using the following equation,

\begin{multline}
\small
E^{form}_{atom}=\frac{1}{20}E^{Nd_2CrFeO_6}_{bulk}-2\times\frac{1}{2}E^{Nd}_{bulk} \\ -\frac{1}{2}E^{Fe}_{bulk}-\frac{1}{2}E^{Cr}_{bulk}-3\times\frac{1}{8}E^{O_2}_{bulk}.
\end{multline}

\noindent The calculated value of the formation energy, E$^{form}_{atom}$, is found -1.4 eV per atom. The negative value of E$^{form}_{atom}$ indicates that this compound is formed by an exothermic process that needs less energy than the breaking of chemical bonds. \cite{Emery2017} To evaluate the thermal stability of Nd$_2$CrFeO$_6$ double perovskite, the phase decomposition energies have been calculated for the two most possible paths by 
\[ Nd_2CrFeO_6\xrightarrow{E_{DP1}} NdFeO_3 + NdCrO_3. \]
\[ Nd_2CrFeO_6\xrightarrow{E_{DP2}} Nd_2O_3 + \frac{1}{2}Fe_2O_3 + \frac{1}{2}Cr_2O_3.  \]
If the decomposition energy for any pathway (E$_{DP}$) is negative, it is not possible to observe a stable region in the phase diagram. \cite{1.5027414}
The estimated decomposition energies for Nd$_2$CrFeO$_6$ double perovskite in both pathways are found to be 2.52 eV f.u.$^{-1}$ ($E_{DP1}$) and 0.86 eV f.u.$^{-1}$ ($E_{DP2}$) suggesting that this compound will not be decomposed spontaneously. This is implying that the requirements for the  thermodynamic stability of Nd$_2$CrFeO$_6$ double perovskite have been satisfied.

To investigate the mechanical stability, elastic constants were calculated using the finite strain method.\cite{qi2018experimental} The calculated elastic constants are summarized in Table 2. For monoclinic crystal, the independent elastic stiffness tensor is reduced to thirteen components, C$_{11}$, C$_{22}$, C$_{33}$, C$_{44}$, C$_{55}$, C$_{66}$, C$_{12}$, C$_{13}$. C$_{15}$, C$_{23}$, C$_{25}$, C$_{35}$ and C$_{46}$ in the Voigt notation. \cite{vajeeston2017first} For a mechanically stable structure, C$_{ij}$ has to satisfy Born–Huang criteria. \cite{vajeeston2017first, PhysRevB.90.224104} The mechanical stability criteria is given by:
\[C_{11}> 0, C_{22}> 0, C_{33}> 0, C_{44}> 0,C_{55}> 0, C_{66}> 0\]
\[C_{33}C_{55}-C_{35}^{2}>0,C_{44}C_{66}-C_{46}^{2}>0,C_{22}+C_{33}-2C_{23}>0\]
\[[C_{22}(C_{33}C_{55}-C_{35}^{2})+2C_{23}C_{25}C_{35}-C_{23}^{2}C_{55}-C_{25}^{2}C_{33})]>0\]
\[2[C_{15}C_{25}(C_{33}C_{12}-C_{13}C_{23})+C_{15}C_{35}(C_{22}C_{13}-C_{12}C_{23})\]
\[+C_{25}C_{35}(C_{11}C_{23}-C_{12}C_{13})]-h +C_{55}g>0\]
where, 

\[g=C_{11}C_{22}C_{33}-C_{11}C_{23}^{2}-C_{22}C_{13}^{2}-C_{33}C_{12}^{2}+2C_{12}C_{13}C_{23}\]

    \small
\[h=C_{15}^{2}(C_{22}C_{33}-C_{23}^{2})+C_{25}^{2}(C_{11}C_{33}-C_{13}^{2}) + C_{35}^{2}(C_{11}C_{22}-C_{12}^{2})\]

\noindent The Nd$_2$CrFeO$_6$ double perovskite meets all essential stability criteria for the monoclinic system. 

\begin{table}
\small
\centering
\caption{\label{tab:table3} The calculated elastic constants, $C_{ij}$ (GPa) of Nd$_2$CrFeO$_6$ double perovskite.}

\begin{tabular}{lllllll}
\hline
$C_{ij}$ & 1 & 2 & 3 & 4 & 5 & 6 \\
\hline
  1 &  281.7 &   131.8  &  117.0   &   0.0    &  2.0   &   0.09 \\
  2    & 131.8  &  238.8  &  113.9  &    0.0   &   2.0  &   0.0 \\
  3&  117.0 &   113.9  &  275.0  &    0.0  &   -4.0   &   0.0 \\
   4&   0.0   &   0.0   &   0.0  &  101.0  &   0   &   3.0 \\
   5&   2.0  &    2.0  &   -4.0  &   0.0  &   85.0  &   0.0 \\
   6&   0.0  &   0.0  &    0.0   &   3.0  &   0.0  &   81.0 \\
    \hline
\end{tabular}
\end{table}

The phonon dispersion curve of Nd$_2$CrFeO$_6$ double perovskite has shown in Fig. 5 to assess the dynamical stability. The structure may be considered dynamically stable if there are no imaginary phonon frequencies in the whole Brillouin zone (BZ). \cite{cheng2014computational,chu2015structural} In this case, we did not observe any imaginary phonon frequency in the whole BZ which indicates that the structure is dynamically stable. 

\begin{figure}[htb]
\centering
\includegraphics[width=8cm]{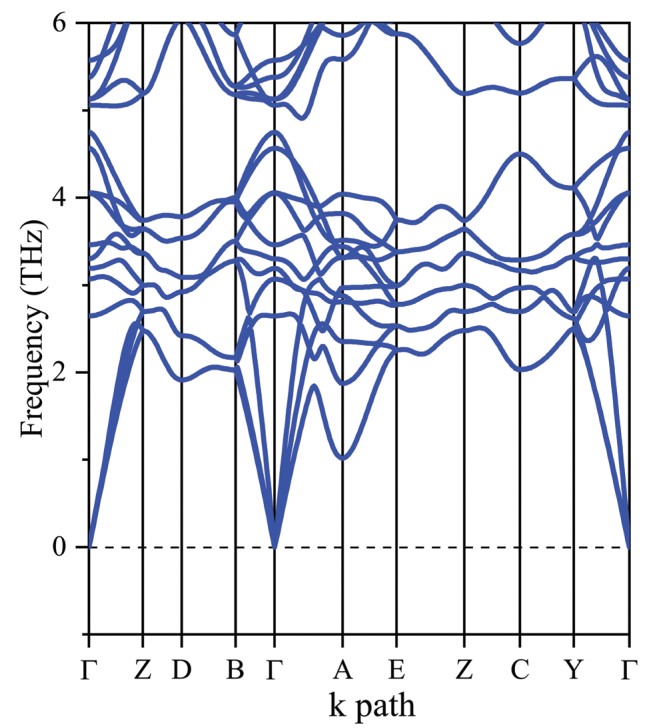}
\caption{\label{fig:epsart} The phonon dispersion curves of Nd$_2$CrFeO$_6$ double perovskite.}
\end{figure}

\section{Magnetic properties}

Table 3 shows the ground-state energy comparison of different spin arrangements (as shown in Fig. 2) relative to the FM spin arrangement and magnetic moment per atom. As discussed above Nd$_2$CrFeO$_6$ double perovskite and its parent single perovskites NdCrO$_3$ and NdFeO$_3$ have G-type AFM order at ground-state. Due to the G-type AFM spin arrangement with Cr$_\uparrow^{3+}$-O-Cr$_\downarrow^{3+}$ and Fe$_\uparrow^{3+}$-O-Fe$_\downarrow^{3+}$ super-exchange interactions in NdCrO$_3$ and NdFeO$_3$ respectively, they do not possess any net magnetic moment. However, Nd$_2$CrFeO$_6$ is quite interesting than NdCrO$_3$ and NdFeO$_3$ as it possess a net magnetic moment of $\sim$2 $\mu_B$ f.u.$^{-1}$ due to the ordered distribution of Cr and Fe cations mediated by Fe$_\uparrow^{3+}$-O-Cr$_\downarrow^{3+}$ super-exchange interaction.

\begin{figure}[htb]
\centering
\includegraphics[width=8cm]{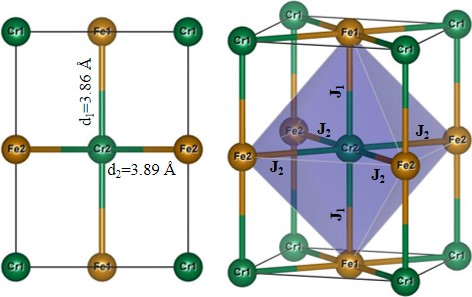}
\caption{\label{fig:epsart} In and out of plane bonds ($d_1$ and $d_2$) in CrFe$_6$ octahedron and magnetic exchange interactions (nearest neighbors, $J_1$ and next nearest neighbors $J_2$) in Nd$_2$CrFeO$_6$ double perovskite}
\end{figure}

\begin{table}[htb]
\centering
\small
  \caption{\ Ground-state energy comparison $\Delta$E (meV/f.u.) of FM, A-AFM, C-AFM and G-AFM states relative to FM state for U$_{Cr, Fe}$=3.50 eV.}
  \label{tbl:example}
  \begin{tabular*}{0.48\textwidth}{@{\extracolsep{\fill}}llllll}
    \hline
    Order & NdCrO$_3$ & NdFeO$_3$ & Nd$_2$CrFeO$_6$ \\
    \hline
    FM    & 0 & 0 & 0\\
    A-AFM & -13 & -86 & -14\\
    C-AFM & -25 & -168 & -27\\
    G-AFM & -36 & -235 & -34\\
    \hline
    M (Cr/Fe)($\mu_B$) & 2.66 & 3.99 & -2.58/4.05\\
    M$_{total}$ ($\mu_B$/f.u.) & 0 & 0 & 2\\
    \hline
  \end{tabular*}
\end{table}

\noindent Furthermore,  the calculated total energies have been used to estimate the magnetic ordering temperature (T$_c$) by mapping to the
classical Heisenberg Hamiltonian:\cite{PhysRevB.64.174402}

\[H=-\sum_{i,j} J_{ij} S_{i}S_{j}\]

\noindent where $\left | S_i \right |=\sqrt{S_i(S_i+1)}$ is the spin at site $i$ and $J_{ij}$ is the spin exchange interaction parameter between $i$ and $j$ sites. Here, we have taken interactions up to the nearest neighbors $J_1$ and second
nearest neighbors $J_2$. It should be noted that two types of bonds (in plane and out of plane) with different lengths are present due to the distortion from the ideal $B^{'}B^{''}_6$ octahedron (as shown in Fig. 6). Therefore, the total energies per magnetic species may be expressed as

\[E_{0}+\sqrt{S_{1}(S_{1}+1)}\sqrt{S_{2}(S_{2}+1)} (2J_{1}+4J_{2})=E_{FM}\]
\[E_{0}+\sqrt{S_{1}(S_{1}+1)}\sqrt{S_{2}(S_{2}+1)} (-2J_{1}+4J_{2})=E_{C-AFM}\]
\[E_{0}+\sqrt{S_{1}(S_{1}+1)}\sqrt{S_{2}(S_{2}-1)} (-2J_{1}-4J_{2})=E_{G-AFM}\]

\noindent where $E_0$ is the spin independent part. The exchange parameters $J_1$=0.62 meV and $J_2$=1.19 meV were calculated by the above equations using $S_1$=5/2 and $S_2$=3/2 for the Fe$^{3+}$ and Cr$^{3+}$ spins. The positive values of exchange interactions are necessarily indicating the AFM nature of Fe-O-Cr interactions. For Nd$_2$CrFeO$_6$, magnetic phase transition temperature has been estimated by the mean-field approximation,\cite{chen2019high}
\[T_c=\frac{2}{3k_B}\sqrt{S_{1}(S_{1}+1)}\sqrt{S_{2}(S_{2}+1)} (2J_{1}+4J_{2})\]

\noindent Since in B-site ordered double perovskite Nd$_2$CrFeO$_6$, the Fe-O-Cr interaction will be dominating, the calculated transition temperature 265 K can be compared with experimentally observed magnetic transition at around 250 K due to Fe-O-Cr interaction in NdFe$_{0.5}$Cr$_{0.5}$O$_3$ perovskite. \cite{SHARANNIA2017109, Wan2016} It should be noted that the classical mean-field approach can give a rough estimation on transition temperature. \cite{Wu2016,Kabiraj2020} To reveal the whole magnetic phase diagram, to understand the microscopic dynamic mechanism and to give a more qualitative picture of the magnetic properties\cite{PhysRevB.104.064431}, further investigation by classical atomistic spin dynamics method based on the DFT calculated exchange parameters is necessary.

\section{Electronic structure}

\begin{figure}[htb]
 \centering
 \includegraphics[width=8.3cm]{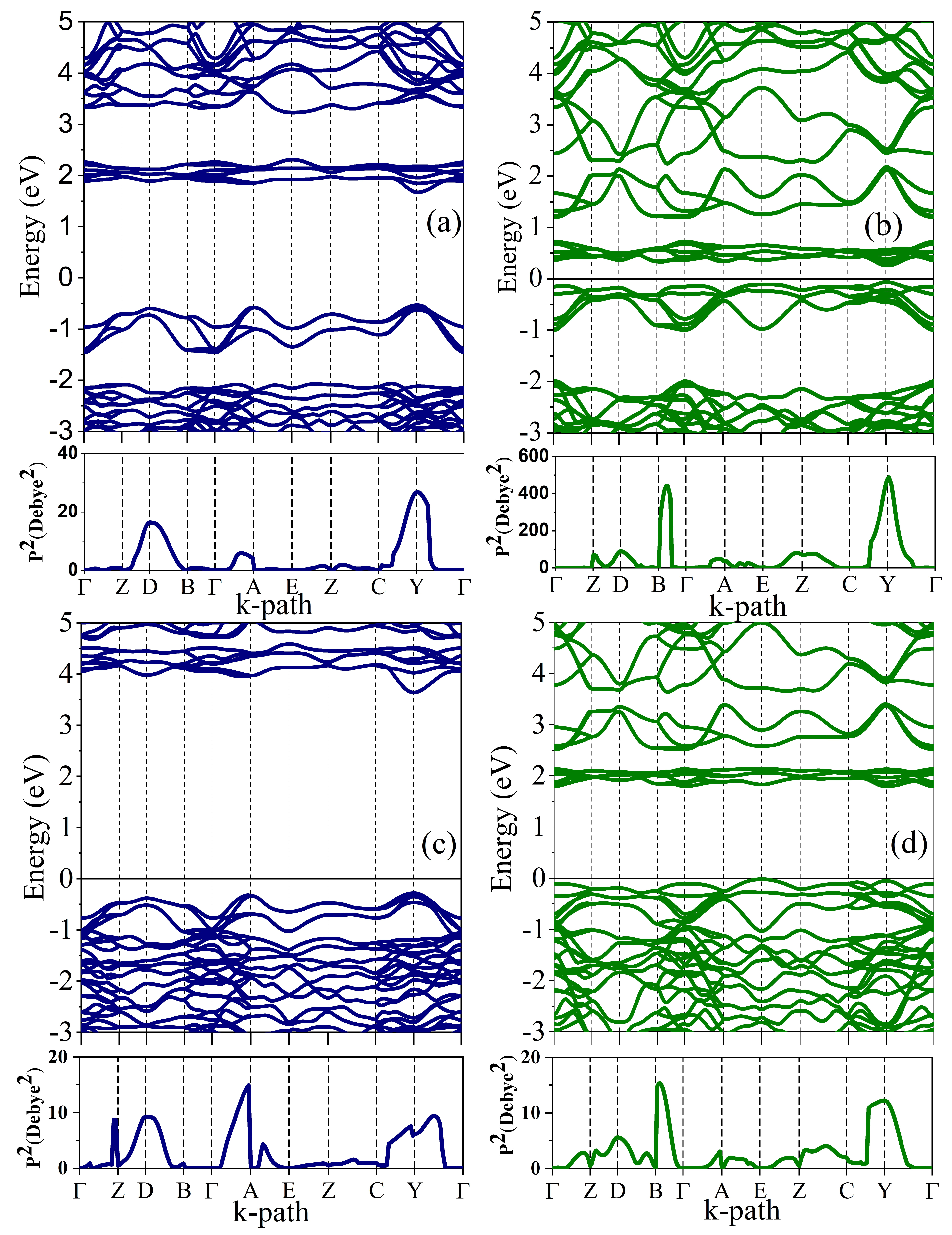}
 \caption{Calculated band structure (top panel) and transition dipole moment (bottom panel) for U$_{Cr}$=U$_{Fe}$=0 in (a) spin-up and (b) spin-down channels. (c) Spin-up and (d) spin-down band structure and transition dipole moment for  U$_{Cr}$=U$_{Fe}$=3.5 eV}
 \label{fgr:example2col}
\end{figure}

\begin{figure}[htb]
 \centering
 \includegraphics[width=8cm]{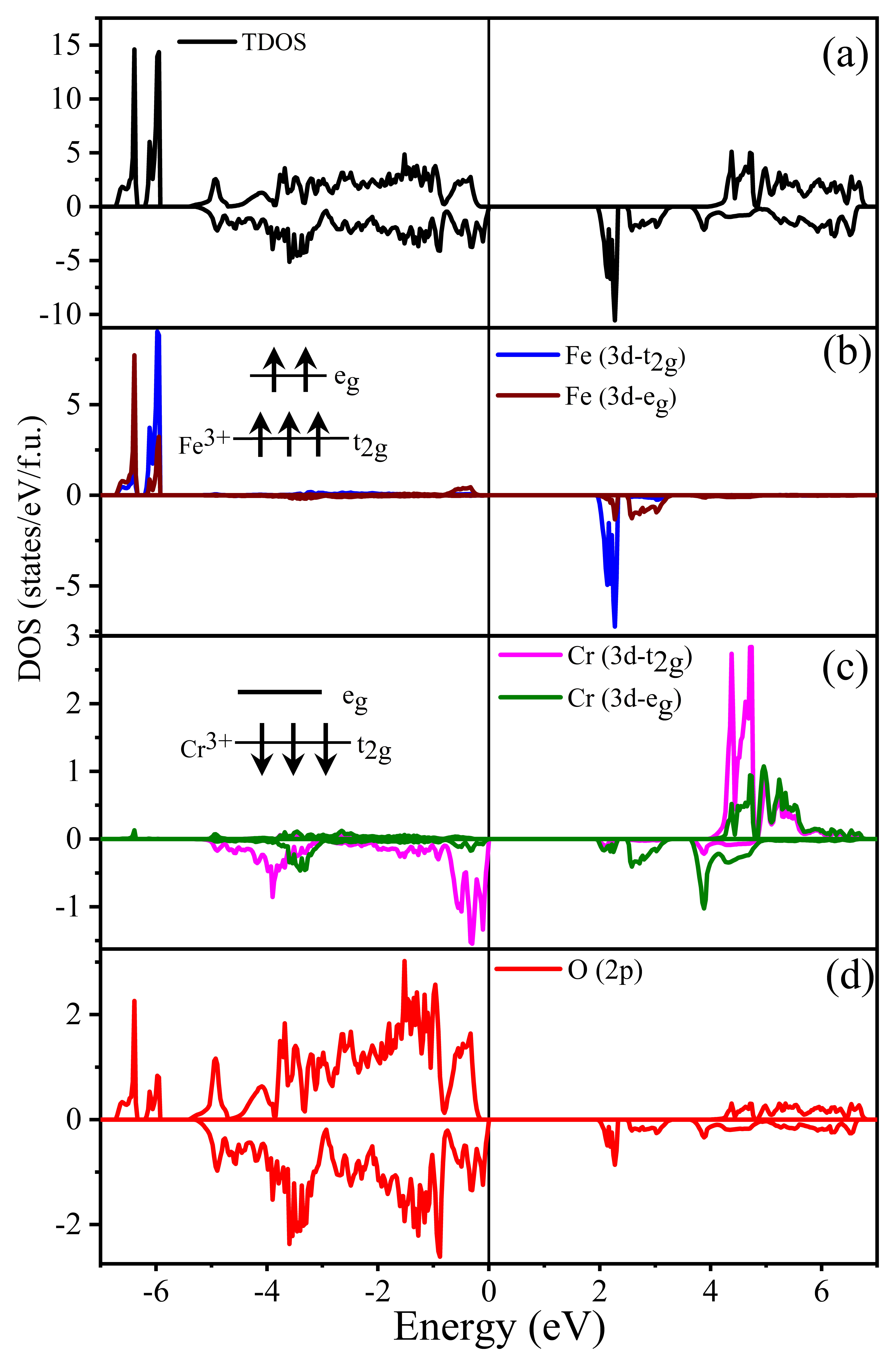}
 \caption{(a) TDOS calculated by GGA+U corresponds to the G-AFM ground state. (b-d) PDOS of Fe (3d), Cr (3d) and O (2p) orbitals respectively. Positive and negative DOS values stand for the up and down spin channel. Fermi level is set at 0 eV.}
 \label{fgr:example2col}
\end{figure}

\begin{figure}[htb]
 \centering
 \includegraphics[width=8.3cm]{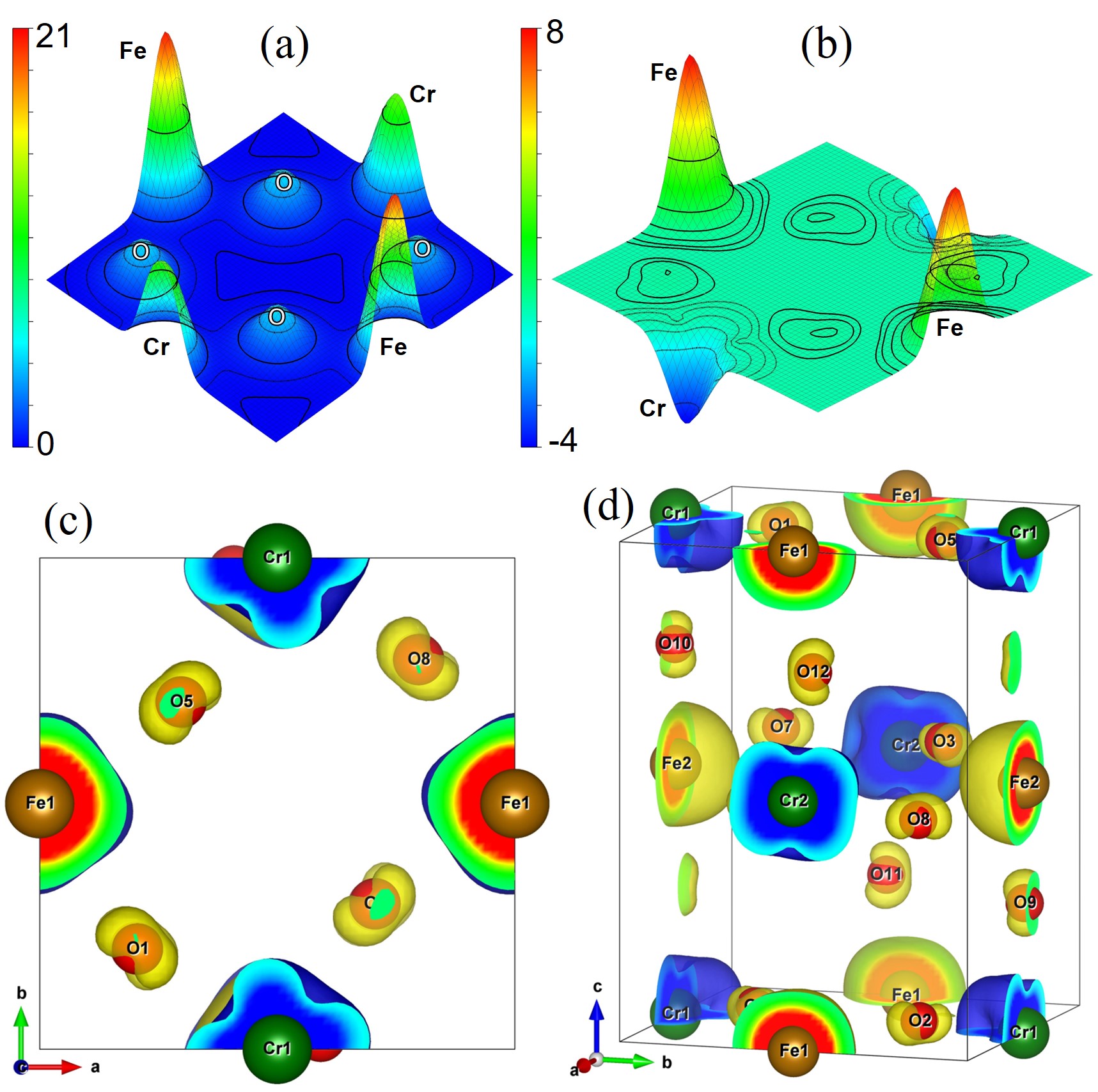}
 \caption{(a) Total charge and (b) magnetization density plots. (c) Cross-sectional and (d) perspective view of three dimensional (3D) magnetization density iso-surface plot at the same iso-value of $\pm$0.01 eV/{\AA$^3$}. The light yellow and light blue colors represent the spin up and spin down states, respectively.}
 \label{fgr:example2col}
\end{figure}

\begin{figure}[htb]
 \centering
 \includegraphics[width=8.5cm]{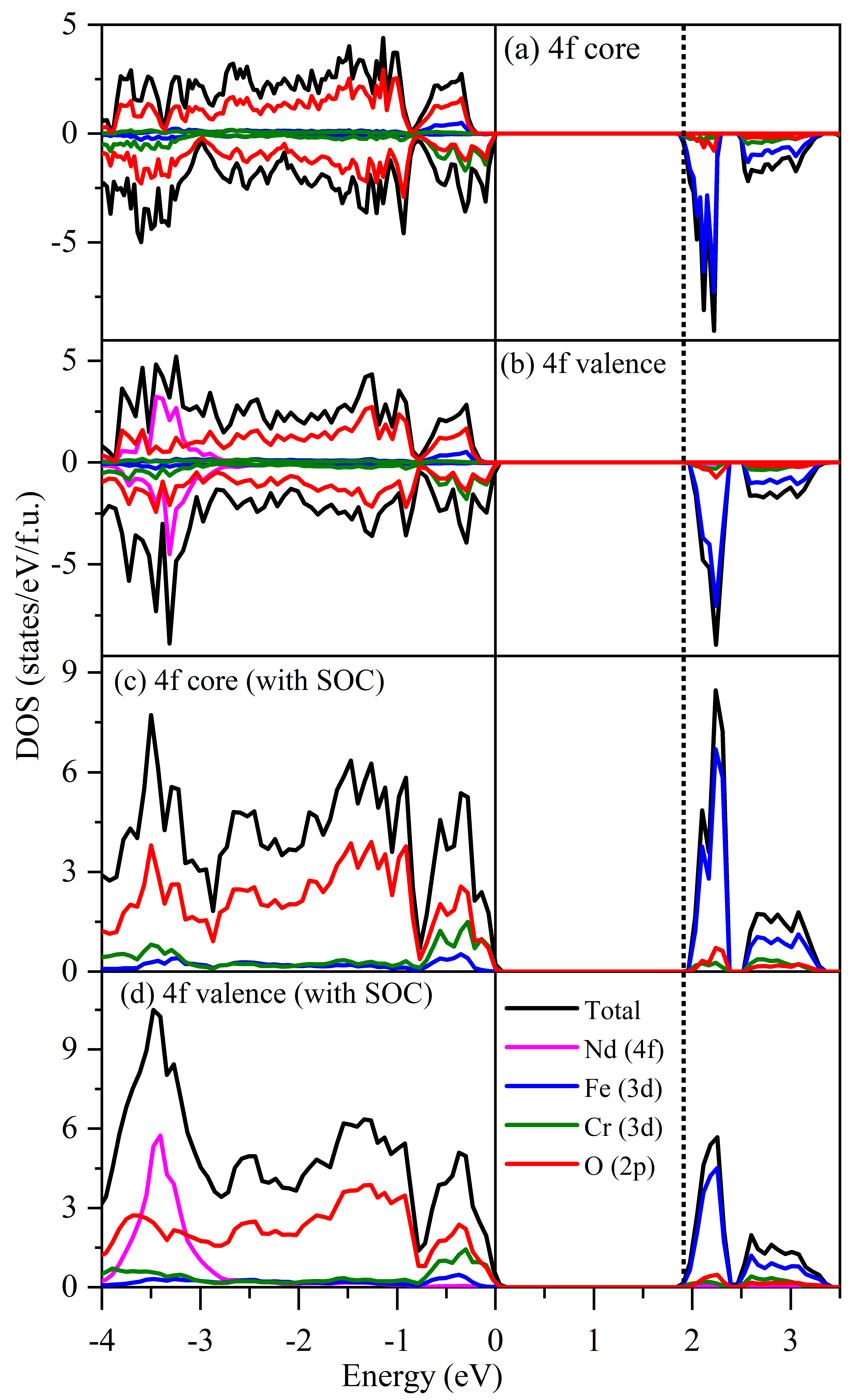}
 \caption{Effect of Nd's 4f electrons and SOC on the electronic structure of Nd$_2$CrFeO$_6$. The calculated PDOS by considering; (a) 4f electron as core electron and (b) 4f electron as valence electron. The obtained PDOS including SOC by considering; (c) 4f electron as core electron and (d) 4f electron as valence electron }
 \label{fgr:example2col}
\end{figure}

Fig. 7 shows spin polarized band structures of Nd$_2$CrFeO$_6$ for \textit{U}$_{Cr}$=\textit{U}$_{Fe}$=0 and 3.5 eV, along the high symmetry directions in the first Brillouin zone for both spin up and spin down channels. The Fermi level (set to 0 eV) was shown by the horizontal line between the valence and conduction bands. The insulator nature of Nd$_2$CrFeO$_6$ has been predicted in both spin channels. For \textit{U}=0 eV (Fig. 7(a) and 7(b)), VBM and CBM are located at the same symmetry point ($Y$) in both spin up and down channels, indicating direct band gap of 2.19 eV and 0.32 eV, respectively. On the other hand, for \textit{U}$_{Cr}$=\textit{U}$_{Fe}$=3.5 eV, in spin down channel, VBM is found to be shifted to the $E$ points, mainly indicating the indirect nature with a 1.82 eV gap. Notably, it can be seen that in down spin channel, VBM at $Y$ point is located just 0.03 eV below the $E$ point. Therefore, in Nd$_2$CrFeO$_6$, the indirect band gap of 1.82 eV ($E\xrightarrow[]{}Y$) and a direct band gap of $\sim$ 1.85 eV ($Y\xrightarrow[]{}Y$) is comparable. The probability of transition either from $E$ or $Y$ points can be understood by the calculated square of the transition dipole moment (\textit{P}$^2$, transition probabilities between two states \cite{WANG2021108033}) and is shown in the bottom panel of Fig. 7(d). At the $E$ point the transition probability is near to zero, whereas a strong probability can be seen around $Y$ point. Therefore, for double perovskite Nd$_2$CrFeO$_6$, the optical absorption could be initiated by direct $Y\xrightarrow[]{}Y$ transition.

To understand the features of VBM and CBM, the total density of states (TDOS) and atom projected orbital decomposed density of states (PDOS) have been analyzed. Fig. 8(a)-(d) display the TDOS of Nd$_2$CrFeO$_6$ double perovskite and PDOS of Fe (3d) Cr (3d) and O (2p) orbitals, respectively. The Fe (3d) and Cr (3d) orbitals are found to be split into $t_{2g}$
and $e_g$ levels due to the FeO$_6$ and CrO$_6$ octahedral environment.
It should be noted that following the local rotation of Fe/CrO$_6$ octahedra in the unit cell orientation, the d$_{{x^2}-{y^2}}$, d$_{yz}$ and d$_{xz}$ orbitals belong to the $t_{2g}$ states. On the other hand, d$_{z^2}$ and d$_{xy}$ orbitals belong to $e_g$ states. \cite{ong2008origin} In the valence band, the DOS in the range from -1 eV to 0 eV in the spin-up channel is dominated by O (2p) orbitals, whereas in the spin-down channel, the VBM is made up of the hybridization of $t_{2g}$ of Cr (3d) and O (2p) states. The CBM was constructed by the hybridization of unoccupied Fe $t_{2g}$ and O (2p) orbitals. From Fig. 8(b), it is evident that both $t_{2g}$ and $e_g$ orbitals of Fe (3d) are filled by electrons in the spin-up channel, while in the spin-down channel $t_{2g}$ and $e_g$ states are completely empty. This demonstrates Fe$^{3+}$($t^3_{2g\uparrow} e^2_{g\uparrow}$) valence state \cite{das2008electronic} in Nd$_2$CrFeO$_6$ as illustrated in the inset of Fig. 8(b). In case of Cr (3d), both $t_{2g}$ and $e_g$ orbitals are found to be empty in spin-up channel and in spin-down channel only the $t_{2g}$ state is occupied, suggesting Cr$^{3+}$($t^3_{2g\downarrow}e^0_{g\downarrow}$) configuration (inset of Fig. 8(c)). These results indicate the strong AFM superexchange between  Cr$^{3+}$($t^3_{2g\downarrow}e^0_{g\downarrow}$) and Fe$^{3+}$($t^3_{2g\uparrow}e^2_{g\uparrow}$) cations \textit{via} O$^{2-}$ anions. For a deeper understanding of the bonding as well as magnetism, charge density analysis have been carried out. Fig. 9(a) shows the total charge density distribution plot, which indicates the dominant Fe-O and Cr-O covalent bonding due to the presence of bond charges along the path between Fe--O and Cr--O. The magnetization density (Fig. 9(b)) displays the antiparallel alignment of spins but with unequal magnitudes demonstrating FiM nature of Nd$_2$CrFeO$_6$. Fig. 9(c) and 9(d) show the cross-sectional and perspective view of the 3D iso-surface of magnetization density. In analogy to the DOS, here the spin density is also mainly contributed by Fe (3d) and Cr (3d) orbitals. The spherical shape around Fe atoms are indicating Fe$^{3+}$($t^3_{2g\uparrow}e^2_{g\uparrow}$) with 3d$^5$ orbital character occupied by spin-up electrons.\cite{PhysRevB.102.174401} On the other hand, only $t^3_{2g}$ orbital are visible around Cr atoms are showing Cr$^{3+}$($t^3_{2g\downarrow}$) character.\cite{Nazir2021} In addition, tiny spin densities have also appeared on the O atoms, due to the Fe$_\uparrow^{3+}$-O$^{2-}$-Cr$_\downarrow^{3+}$ superexchange coupling. To investigate the effects of Nd's 4f spin and SOC, further calculations have been carried out considering Nd's 4f electrons as valence electrons with U$_{Nd}$= 6.5 eV. Including Nd's 4f electrons in valence shell, the G-AFM ordering remain as the lowest energy spin state, which is in agreement with the previous reports on ReFeO$_3$ and ReCrO$_3$ perovskites.\cite{zhu2017electronic, Zhao2017} Fig. 10(a) and 10(b) show the comparative DOS plots using 4f electrons as core and valence electrons, respectively. No contribution of Nd (4f) orbital can be seen either in VBM and CBM from Fig. 10(b). Moreover, the band gap change (<0.8 eV) was found to be negligible. Similar results have been found from the calculations with including SOC effect as shown in Figs. 10(c) and 10(d). Therefore, we think that effects of Nd's 4f spin and SOC are not so significant that change our conclusion with the results obtained by considering Nd's 4f electron as core electrons without SOC.

Before concluding, it could be worthwhile to compare the calculated magnetic properties and band gap of Nd$_2$CrFeO$_6$ with other similar B-site ordered double perovskite oxides. Table 4 shows the magnetic order, transition temperature and band gaps of some Re containing B-site ordered double perovskites along with our DFT-calculated results. The obtained ground state is in agreement with Re$_2$CrFeO$_6$ and band gap is within the visible range like other candidates. Therefore, it is anticipated that this compound may be a potential candidate in various intriguing applications including spintronic memory devices, memristive high-performance data storage devices, light-emitting diodes, cryogenic magnetic cooling devices, solar cell and visible light driven photocatalysis. 

\begin{table}[htb]
\centering
\small
\renewcommand{\arraystretch}{1}
  \caption{Magnetic order (M), transition temperature (T$_c$) and optical band gap (E$_g$) of some Re$_2$B$^{'}$B$^{''}$O$_6$ materials.}
  \label{tbl:example}
  \begin{tabular*}{0.48\textwidth}{@{\extracolsep{\fill}}llllll}
    \hline
    Compound & B-site & M & T$_c$ (K) & E$_g$ (eV) & Ref. \\
    \hline
     La$_2$MnCoO$_6$ & ordered & FM & 226 & 1.93 & \cite{gauvin2018electronic,chang2017effect}\\
    La$_2$NiMnO$_6$ & ordered & FM & 270 & 1.42 & \cite{acsaelm.8b00062}\\
    Nd$_2$NiMnO$_6$ & ordered & FM & 195 & 1.57 & \cite{acsaelm.8b00062}\\
    Dy$_2$NiMnO$_6$ & ordered & FM & 105 & 1.62 & \cite{chanda2015structural}\\
    La$_2$FeCrO$_6$ & ordered & FiM  & 45 & 1.6 &\cite{ohtomo2012spontaneous} \\
    Pr$_2$FeCrO$_6$ & partial & FiM & 245 & 2.13 & \cite{gaikwad2019structural}\\
    Nd$_2$CrFeO$_6$ & ordered & FiM & 265 & 1.85 & This work\\
    \hline
  \end{tabular*}
\end{table}

\section{Conclusions}

This article focused on the phase stability, magnetic properties,
and electronic band structure of B-site ordered Nd$_2$CrFeO$_6$ double perovskite based on DFT calculations. The value of on-site Coulomb interaction \textit{U} was chosen to be 3.5 eV for 3d-electrons of Cr and Fe by  applying a range of \textit{U} values within GGA+\textit{U} calculations that can reproduce the band gaps and magnetic ground state of parent single perovskites NdCrO$_3$ and NdCrO$_3$ as well as double perovskite Nd$_2$CrFeO$_6$. The material has been predicted to be stable by means of thermal, mechanical, and dynamic stability criteria. Compared to the AFM nature of the parent single perovskites counterparts, B site ordered double perovskite Nd$_2$CrFeO$_6$ demonstrates a ferrimagnetic ground-state with a total magnetization of 2 $\mu_B$ per formula unit due to the Fe$_\uparrow^{3+}$-O-Cr$_\downarrow^{3+}$ super-exchange interaction. The band structure and density of states calculations confirm the semiconducting nature of this material with a direct band gap of $\sim$1.85 eV which is significantly smaller than the band gaps of its parent single perovskites. Notably, the probability of direct transition from VBM to CBM has been found to be higher by using both GGA and GGA+\textit{U} approaches, which is essential for many optical applications such as solar cell, LASER \textit{etc}. Due to ferrimagnetism along with the direct nature of optical band gap, B-site ordered Nd$_2$CrFeO$_6$ double perovskite could be a potential candidate for the spintronic and visible light-mediated optical applications. Notably, the crystallographic structure and physical properties of most of the single perovskite materials are experimentally well established, however, double perovskites are comparatively less investigated. Therefore, the experimental data for various double perovskites are always not available. Our present investigation might pave the way for the theoretical prediction of the physical properties of any new double perovskite materials by employing the experimental parameters of their parent single perovskites and optimizing the effects of on-site d-d Coulomb interaction energy within GGA+\textit{U} calculations.  

\section*{Conflicts of interest}
``There are no conflicts to declare''.

\section*{Acknowledgements}
The computational facility provided by 'Ogata Laboratory (theoretical solid mechanics), Department of Mechanical Science and Bioengineering, Osaka University, Japan' is sincerely acknowledged. The authors would also like to acknowledge the Committee for Advanced Studies and Research (CASR), BUET for financial assistance.


\bibliography{rsc} 
\bibliographystyle{rsc} 


\end{document}